\documentclass[10pt,b5paper,twoside]{ise2}

\usepackage{graphicx}
\usepackage{natbib}

\begin{document}
\sloppy

\def\nuInu{$\nu$I$_{\nu}$}
\def\dnuInu{$\delta$($\nu$I$_\nu$)}
\def\nWmsr{nWm$^{-2}$sr$^{-1}$}

\title{The Breakout of Protostellar Winds in the Infalling Environment}
\author{Francis P. Wilkin \\
\institutename{Instituto de Astronom\'{\i}a}, UNAM \\
\instituteaddress{Apdo. Postal 3-72 (Xangari), 
58089 Morelia, Michoac\'an, Mexico} \\
\email{\inst{1}f.wilkin@astrosmo.unam.mx@astrosmo.unam.mx}}

\date{}

\maketitle

\begin{abstract}
\noindent
The time of protostellar wind breakout may be determined by the
evolution of the infalling flow, rather than any sudden
change in the central engine. I examine the transition from pure infall 
to outflow, in the context of the inside-out collapse of a rotating 
molecular cloud core. I have followed numerically the motion of 
the shocked shell created by  the impact of 
a stellar wind and infalling gas. 
These fully time-dependent calculations  include cases both where 
the shell falls back to the stellar surface,  
and where it breaks out as a true outflow. 
Assuming a wind launched from the protostellar surface, the
breakout time is determined in terms of the parameters describing the 
wind (${\dot M}_w$, $V_w$) and collapsing cloud core 
($a_\circ$, $\Omega$). 
The trapped wind phase consists of a wind sufficiently
strong to push material back from the stellar surface, but too weak
to carry the heavy, shocked infall out of the star's gravitational 
potential. To produce a large-scale outflow, 
the shocked material must be able to climb out of the star's gravitational 
potential well, carrying with it the dense, swept-up infall. 
\\

\end{abstract}

\section{Introduction}
Because essentially all known protostellar or pre-main sequence objects
show evidence of winds, jets, or outflows, the current thinking is
concentrated on a picture of simultaneous infall and outflow, in
which infall and accretion occur towards the protostellar 
equatorial regions, and a wind breaks out along the poles 
(e.g. \citet{awtb}).  Yet early  thinking on the stages 
of young stellar evolution identified a phase 
in which infall directly strikes the protostellar surface, 
with no outflow present \citep{sal}. 
No clear examples are known of such protostars. 
In this contribution  I consider limits on the timescale 
for purely-accreting  objects in  the context of 
the standard  model of inside-out collapse 
from a molecular  cloud core. 
The mathematical formulation is a generalization of \citet{wsI}, 
dropping the assumptions 
of normal force balance and quasi-stationarity to permit 
full time-dependence and dynamical expansion (or collapse).

\section{Description of the Infall, Wind, and Protostar}
The inside-out collapse of a singular, 
isothermal sphere yields a mass accretion rate 
${\dot M}_i\,=\,0.975\,a_\circ^3/G$ at
the center \citep{shu77}. Here $a_\circ$ is the isothermal 
sound speed in the cloud core. At the origin is 
a protostar whose mass grows 
linearly in time $M_*={\dot M}_i\,t$, 
where $t$ is the time since the start of collapse. In the presence
of initial, solid-body rotation, the infall is distorted, and accretion
occurs preferentially onto the circumstellar disk 
\citep{ulrich, cm, tsc}. 
The natural length scale
of the distortion is the centrifugal radius $R_{cen}$, 
which grows as $t^3$. I  turn on a wind at the stellar surface, of radius 
$R_*$, and numerically determine whether 
it can halt infall and escape. At early times, $R_{cen}\,\ll\,R_*$ 
and the accretion is nearly isotropic, 
making breakout of the wind difficult.  
At late times, when $R_{cen}\,\gg\,R_*$, 
escape becomes easy along the poles.

For simplicity, the wind is assumed isotropic and of constant speed $V_w$, and
mass-loss rate ${\dot M}_w$. The wind and infall collide supersonically,
and a shocked shell forms. Low speeds imply rapid cooling and a 
geometrically thin shell. The dynamics of such time-dependent, 
thin shells has been discussed in detail by \citet{giu}. 
I include the inputs of mass and momentum from
infall, the wind, and the gravitational force due to the protostar. 

\begin{figure}[h!]
	\centerline{\resizebox{110mm}{!}{\includegraphics{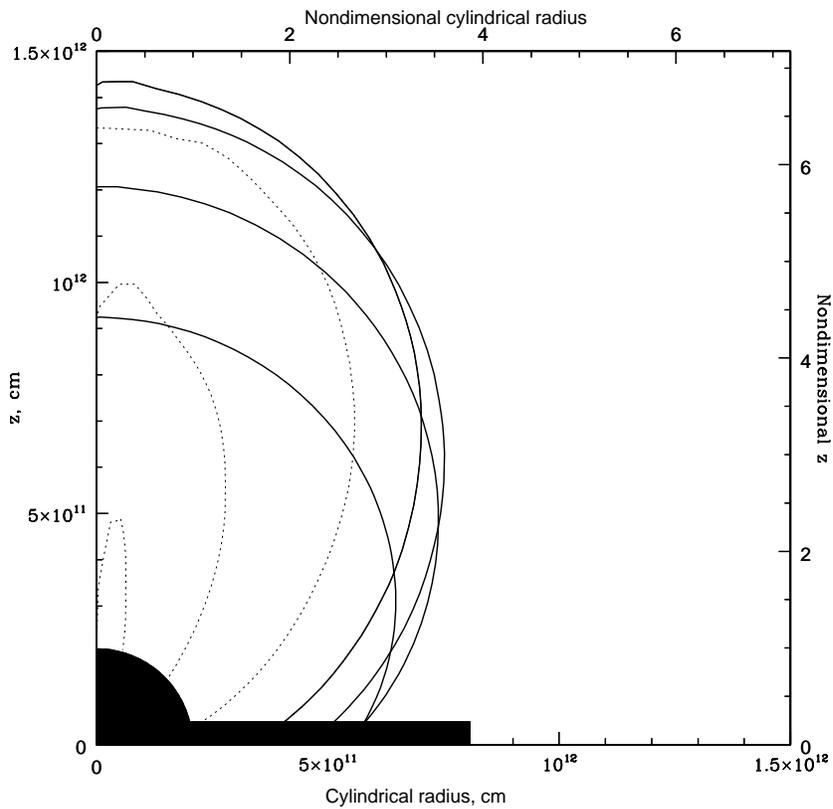}}}	
\caption[]{\small Time-evolution of a failed outflow.
The solid curves show the initial rise, while dotted curves 
show the subsequent recollapse. 
The protostellar age since core formation 
is  $3.8\times\,10^4$ years. 
The shapes correspond to equal time intervals of $0.016$ years.
The size of the centrifugal radius is indicated by the disk. 
Lengths are in cm on the left and bottom axes, 
and in units of the stellar radius on the remaining axes.}
\label{fig:riseandfall}
\end{figure}

\section{The Trapped Wind Stage}
\begin{figure}[h!]
	\centerline{\resizebox{110mm}{!}{\includegraphics{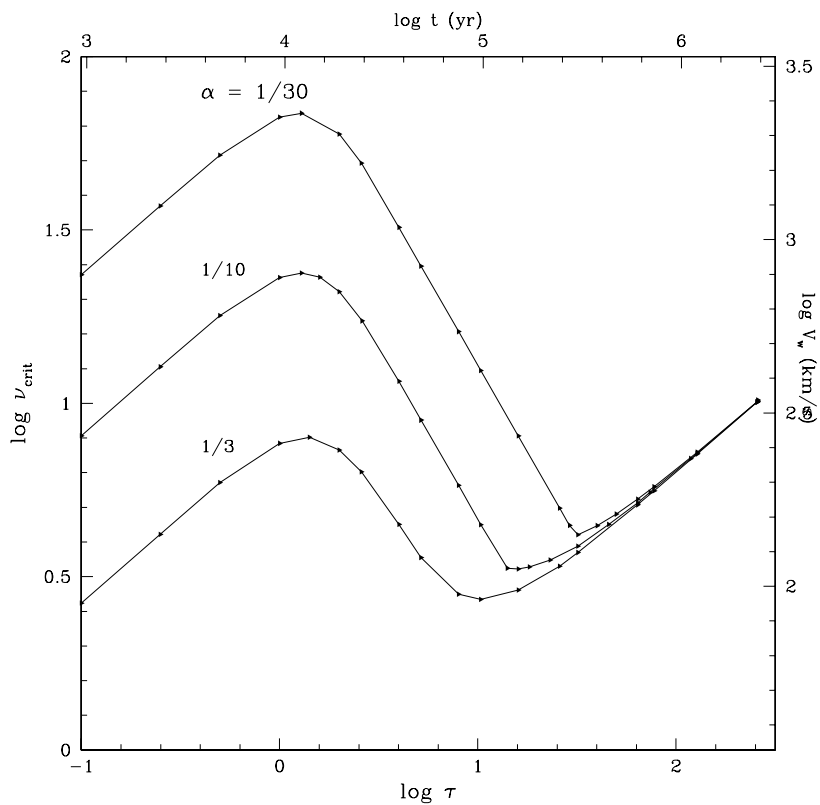}}}	
\caption[]{\small Minimum breakout wind speed versus evolutionary time.
The three loci correspond to three ratios, $\alpha$, of the 
wind mass loss to infall accretion rate. For a given $\alpha$, 
the region above the curve corresponds to breakout, while that below
the curve corresponds to recollapse. 
The power-law increase in $\nu_{crit}$ at early and
late times is associated with the increasing gravitating mass of the
protostar.}
\label{fig:nofit}
\end{figure}

\begin{figure}[h!]
	\centerline{\resizebox{110mm}{!}{\includegraphics{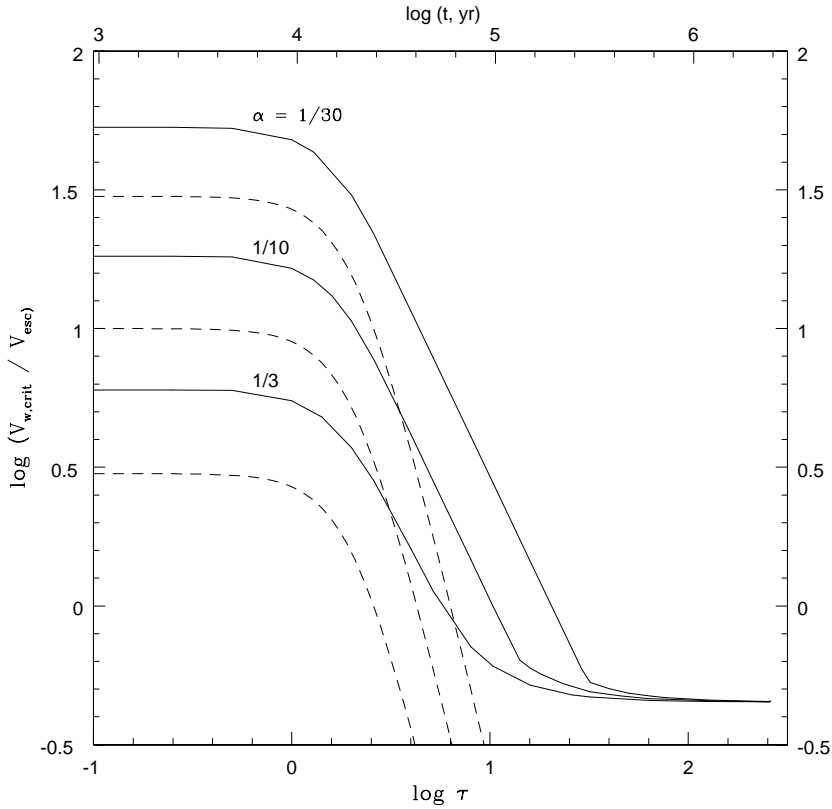}}}	
\caption[]{\small Critical wind speed for breakout (solid curves),  in units 
of the free-fall (escape) speed, as a function of evolutionary time. 
The corresponding $\alpha$-values
are shown, as well as the wind speed necessary for ram pressure balance
at the stellar surface (dashed curves). Assuming wind launch conditions
$V_w/V_{esc}=const.$ (i.e. following a horizontal line in this 
figure), evolution begins at the left edge of the plot with
the wind unable to advance beyond the stellar surface until 
the line intersects the appropriate dashed curve. Then 
the ``trapped wind'' phase lasts until the
line intersects the corresponding solid curve for breakout. 
For example, for $V_w/V_{esc}=1.6,$ we follow a horizontal line at 
$\log(V_w/V_{esc})=0.2$. 
The trapped wind phase begins at $t\approx 19,000\,yr$,
while breakout occurs only at $t\approx 38,000\, yr$, 
indicating a substantial duration for the trapped wind stage.}
\label{fig:nonsteady}
\end{figure}

For a given evolutionary time and 
ratio $\alpha\equiv{\dot M}_w/{\dot M}_i$, 
I determine the minimum wind 
speed necessary to break out of the infalling flow. 
Figure~\ref{fig:riseandfall} shows an example calculation of a shell 
that fails to escape and falls back to the star. 
In this case, a modest increase in wind speed 
dramatically changes the outcome, permitting breakout
of the shell from the infall region. 
I solved the problem in dimensionless form, 
which reduces the parameter space from 
six ($R_*,\,{\dot M}_w,\,V_w,\,a_\circ,\,\Omega,\,t$) 
to three dimensions (nondimensional time $\tau$, 
wind speed $\nu$, and $\alpha$). 
As a result, the parameter space has been fully explored.
Figure~\ref{fig:nofit} shows 
the critical wind speed for breakout. 
When the wind ram pressure exceeds the infall ram pressure at
the stellar surface, the wind may initially push the shell 
upwards. But if the wind speed is less than the critical speed,
the shell stalls and falls back. This is the trapped wind stage. 
In Figure~\ref{fig:nonsteady}, 
its duration is the time between the dashed 
(ram pressure balance) curve and the solid one (critical wind speed) 
for a given ratio $\alpha\equiv {\dot M}_w/{\dot M}_i$. 


\section{Discussion}
The results of Figures~\ref{fig:nofit} and~\ref{fig:nonsteady} 
may  be easily scaled to apply to {\it anisotropic} winds, 
by comparing to an equivalent, 
isotropic wind having the same mass and momentum loss rates along
the z-axis.  Indeed, at early times, 
the evolution is primarily determined 
by the momentum loss rate of the wind in this direction. 
At late times, it is the wind speed rather than momentum loss rate
that determines breakout. It is hoped that these semi-analytic results 
will inspire more detailed exploration of this problem with fully
radiative hydrodynamic simulations. The existing literature 
(e.g.~\citet{frank1994}) on this is not immediately comparable 
because they have used a different density law which is self-similar, 
unlike that of Cassen \& Moosman. Moreover, I argue that initial 
conditions with wind velocity much greater than the critical velocity
are unphysical, as such a strong wind would have broken out at an earlier 
time, unless the wind itself evolves strongly with time. 

Strong collimation of wind by anisotropic 
infall is not seen in the current calculations, 
although further exploration of this issue is
forthcoming. I note that numerical simulations demonstrating
strong collimation due to the circumstellar density 
asymmetry (e.g.~\citet{delamarter}) have assumed a much more 
asymmetric density field than that used here.

\begin{acknowledgement}
I am grateful to the Observatoire de la C\^ote d'Azur for a Henri Poincar\'e 
Fellowship and to the D\'epartement Fresnel, 
UMRS 5528 for hosting my postdoctoral stay in France.   
I also thank CONACyT/M\'exico for financial support and S.Stahler 
for encouragement in this work. 
\end{acknowledgement}

\end{document}